\documentclass{ws-procs975x65}
\begin{document}

%\hfill Preprint numbers: ITP-UU-11/04, SPIN-11/02
%FER-xx-xx
%arXiv:1101.5059 [gr-qc], http://arxiv.org/abs/arXiv:1101.5059
%\leftline{\hfill Date: \today}

%\vspace{0.5cm}

\title{Boson stars with nonminimal coupling}

\author{Anja Marunovi\'c}

%$^a$\footnote{a.marunovic@uu.nl},
%Dubravko Horvat$^b$~\footnote{dubravko.horvat@fer.hr} and Miljenko
%Murkovi\'c$^c$~\footnote{mmurkovi@kbc-zagreb.hr}}

\address{Institute for Theoretical Physics, Spinoza
Institute and
 Center for Extreme Matter and Emergent Phenomena,
 Utrecht University, Postbus 80.195,
  3508 TD Utrecht, The Netherlands\\
Email: a.marunovic@uu.nl }

\begin{abstract} \noindent
Boson stars coupled to Einstein's general relativity possess some
features similar to gravastars, such as the anisotropy in principal
pressures and relatively large compactness ($\mu_{max} = 0.32$).
However, no matter how large the self-interaction is, the ordinary
boson star cannot obtain arbitrarily large compression and as such
does not represent a good black hole mimicker. When the boson star
is nonminimally coupled to gravity, the resulting configurations
resemble more the dark energy stars then the ordinary boson stars,
with compactness significantly larger then that in ordinary boson
stars (if matter is not constrained with the energy conditions). The
gravitationally bound system of a boson star and a global monopole
represents a good black hole mimicker.
\end{abstract}

%PACS:
%04.62.+v Quantum fields in curved spacetime
%04.70.Dy Quantum aspects of black holes, evaporation, thermodynamics

%\maketitle
\bodymatter

\section{Introduction}
\label{Introduction}
If matter obeys the strong energy condition (SEC), classical general relativity predicts the existence of black holes. (According to
the SEC the sum of the energy density and pressures, $\rho +p_i \ge 0$ , cannot be
negative.) Black holes are stationary, vacuum solutions of the Einstein equations
that possess an event horizon and a physical singularity which is hidden by the
event horizon. This physical singularity represents an as-yet-unresolved problem,
since it implies the controversial information loss paradox according to which any
information will get completely lost on the singularity of a black hole. The principles
of information loss are in conflict with the standard laws of quantum physics. Taking all these
considerations into account it is natural to question whether the final stage of a
massive star collapse is a black hole, or perhaps some other as-yet-not-understood
dense object, that prevents further collapse.

Sakharov was the first that introduced the concept of nonsingular collapse
through the equation of state for the cosmological dark energy (for which the
pressure is negative, $p=-\rho$) as a super-dense fluid and then Gliner assumed
that such a fluid could be the final state of the gravitational collapse. Inspired
by these ideas Mazur and Mottola investigated alternative configurations
which led to a solution dubbed the gravastar (gravitational vacuum star)~\cite{MM}.
This anisotropic, highly compact astrophysical object consists of a de Sitter core,
which through a vacuum transition layer matches on an exterior Schwarzschild
space-time by avoiding an event horizon formation. Due to their high compactness
(defined as the ratio of the mass to radius) gravastars are perceived by distant
observers as black holes, and hence they can be good black hole mimickers.

Apart from the de Sitter core, gravastars possess the peculiar
property of pressure anisotropy~\cite{AM1}. Microscopic models of gravastars would shed light on the above mentioned
problems of curvature singularities and black hole information loss
paradox.

\section{Boson stars nonminimally coupled to gravity}

The oldest, and accordingly the most studied, astrophysical example based on
the Lagrangian formalism, is the boson star, which is a compact object built from
a self-interacting, gravitationally bound scalar field. It is known that boson
stars coupled to Einstein's general relativity possess some features that characterize
gravastars, such as the anisotropy in principal pressures and relatively large
compactness ($\mu_{max} = 0.32 $). Here we extend the analysis of boson stars and modify
the Einstein-Hilbert action by
introducing a nonminimal coupling of the scalar field to gravity via the Ricci
curvature scalar~\cite{AM2}~\footnote{The metric is spherically symmetric
$ ds^2=-e^{\nu(r)}dt^2+e^{\lambda(r)}
dr^2+r^2d\theta^2+r^2\sin^2\theta d\varphi^2\,. $}:

\begin{equation} S_\phi=\int d^4 x
\sqrt{-g}\left(-g^{\mu\nu}\partial_\mu\phi^*\partial_\nu\phi-m^2
\phi^*\phi-\frac {\lambda_\phi}{2} (\phi^*\phi)^2+\boxed{\xi R
\phi^*\phi}\right)\,,
\end{equation}
\begin{figure}
\begin{center}
\leavevmode
\includegraphics[scale=0.55]{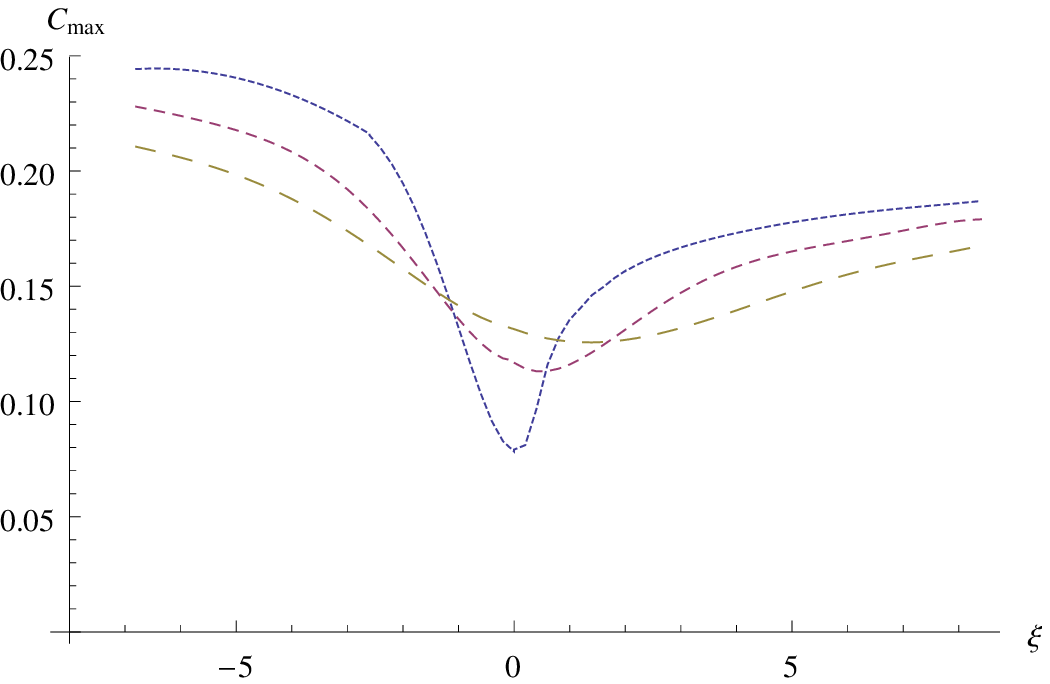}\qquad
\includegraphics[scale=0.55]{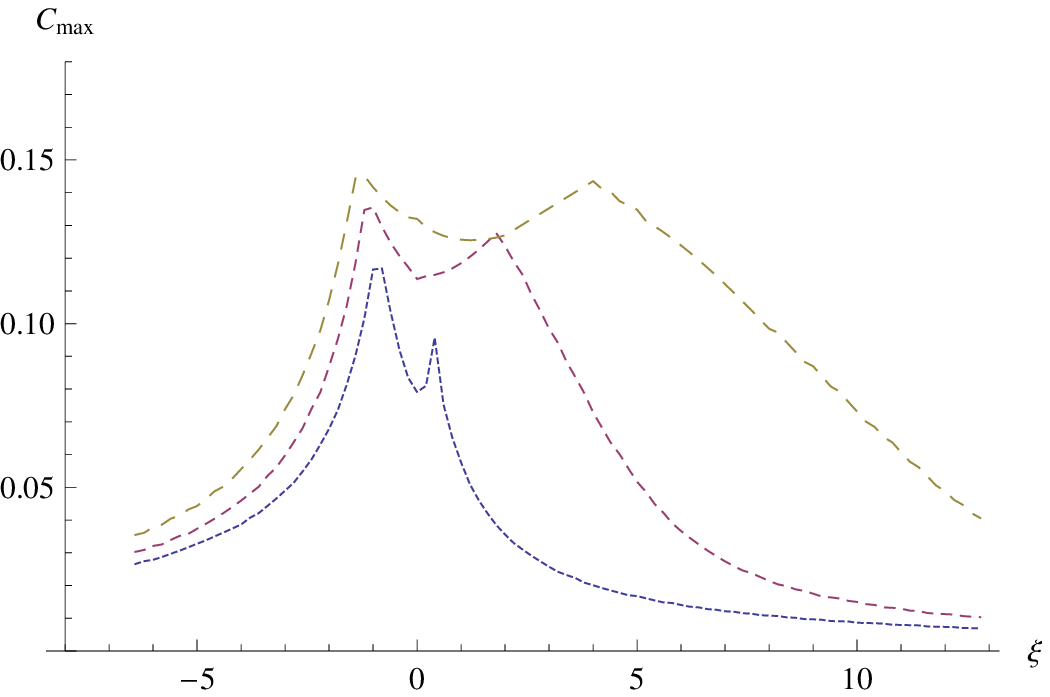}
\end{center}
\caption{Effective compactness as a function of $\xi$ for
$\lambda_\phi=0$ (dotted curve), $\lambda_\phi=20$
(short-dashed) and $\lambda_\phi=50$ (long-dashed). Right: Effective compactness if weak energy condition (WEC) and dominant energy condition (DEC) are obeyed.} \label{figEC}
\end{figure}

\begin{figure}
\leavevmode
\includegraphics[scale=0.55]{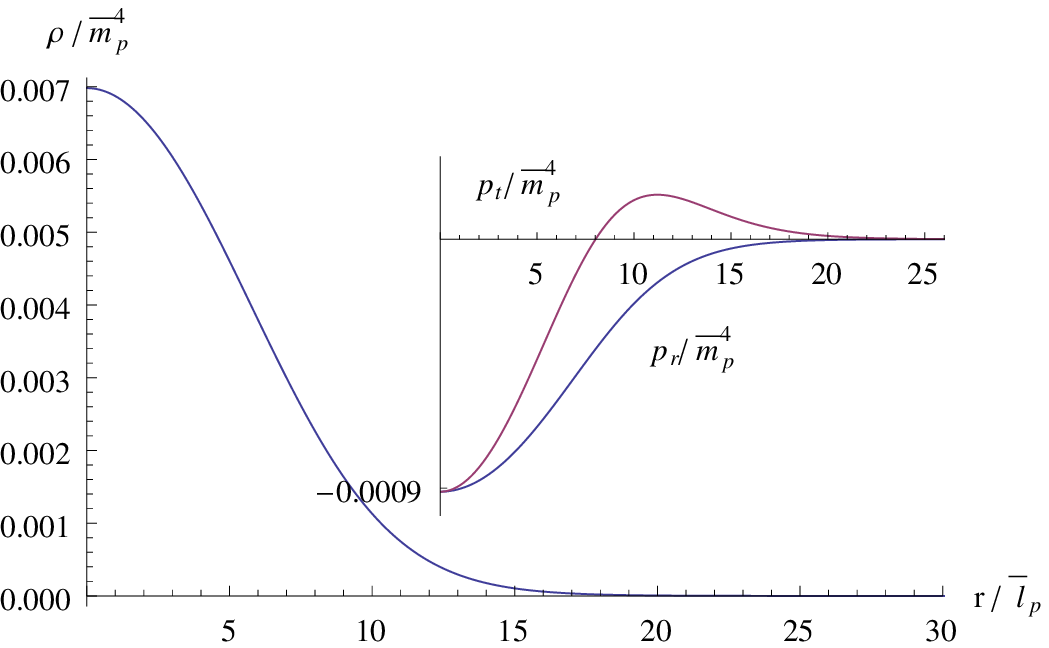}\qquad
\includegraphics[scale=0.55]{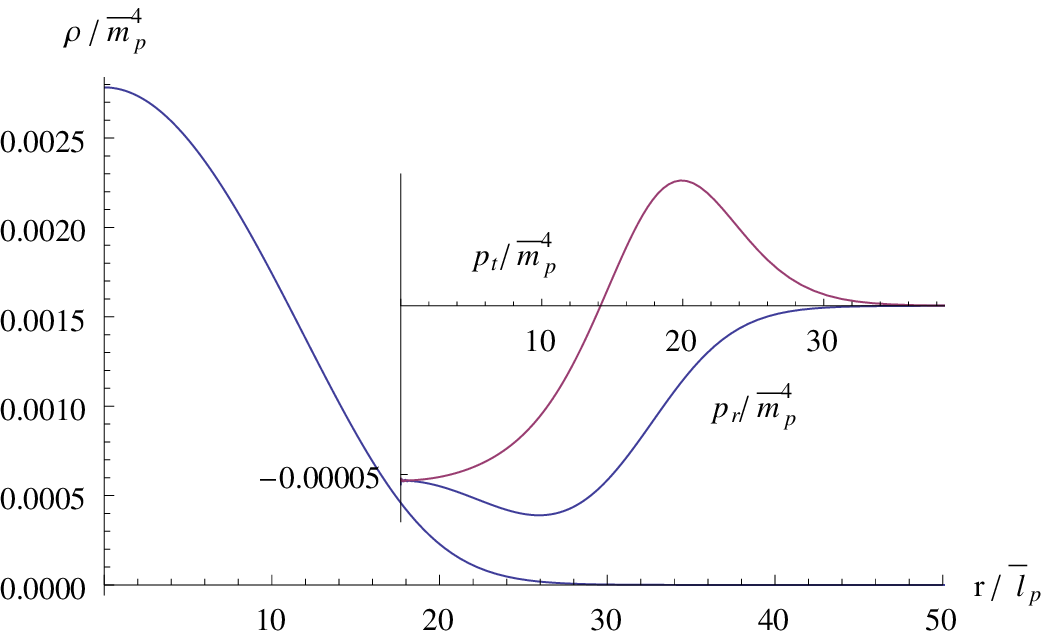}
\caption{Energy density and pressures (inset) for: a) Left: $\xi=-4$,
$\{\lambda_\phi,\sigma_c\}=\{0,0.050\}$; b) Right:
$\xi=-4$, $\{\lambda_\phi,\sigma_c\}=\{100 ,0.034\}$.}
\label{TDKm}
\end{figure}

We show that already this minimal extension of general relativity
results in configurations that resemble more the dark energy stars (due to negative pressures) then the
ordinary boson stars (see Fig.~\ref{TDKm}), with compactness~\footnote{Effective compactness is defined as $C_{eff}=M_{99}/R_{99}$ where $R_{99}$ is the radius at which the mass, defined in terms of the metric function $e^{-\lambda}=1-2m/r$ equals $99\%$ of
the total mass $M = m(\infty)$. Compactness is $\mu=2m/r=2 C_{eff}$. } significantly larger then that in ordinary
boson stars (if matter is not constrained with the energy conditions) (see Fig.\ref{figEC}). It is interesting to observe that, if matter is not constrained with the energy conditions, $\mu_{max}\approx 0.5$ in the region $\xi<0$ where the pressures are negative. However, if matter obeys the energy conditions, $\mu_{max}\approx 0.32$, and hence not much larger then in the case of ordinary boson stars.

\section{Boson defect stars}

Another field-theoretic model that we investigate involves a
 combined system of a boson star and a global monopole~\cite{AM3}.
The simplest field-theoretic realization of the global monopole includes a scalar
field theory with an (global) $O(3)$ - symmetry which is spontaneously broken to
$O(2)$ by the vacuum.

The action that governs the dynamics of a global monopole is:
\begin{equation} S_\phi=\int d^4 x
\sqrt{-g}\left(-\frac 12 g^{\mu\nu}\partial_\mu(\phi^a)\partial_\nu(\phi^a)-V(\phi^a)+\frac 12 \xi R(\phi^a\phi^a)\right)
\end{equation}
and the potential is a standard symmetry breaking Mexican hat potential,
\begin{equation}
V(\phi^a\phi^a)=\frac{\mu^2}{2}\phi^a\phi^a+\frac{\lambda}{4}(\phi^a\phi^a)^2+\frac{\mu^4}{4\lambda}\,. \end{equation}
A global monopole is a solution of the field for the given action, with the so called \emph{hedgehog} Ansatz
%\begin{equation} \left(\phi^a\right)=\phi(r,t)\begin{pmatrix} \sin\theta\cos\varphi %\\\sin\theta\sin\varphi \\\cos\varphi \end{pmatrix}
\begin{equation}
\phi^a=\phi(r)(\sin\theta\cos\varphi, \sin\theta\sin\varphi, \cos\varphi).
\end{equation}

\begin{figure}
\centering
\includegraphics[scale=0.55]{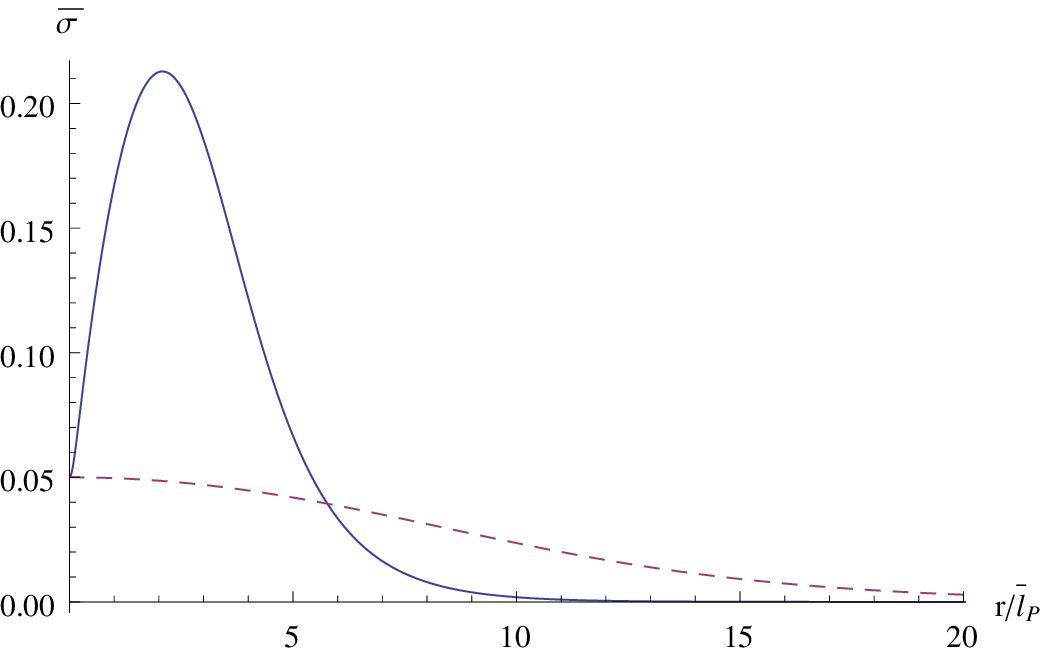}\qquad
\includegraphics[scale=0.55]{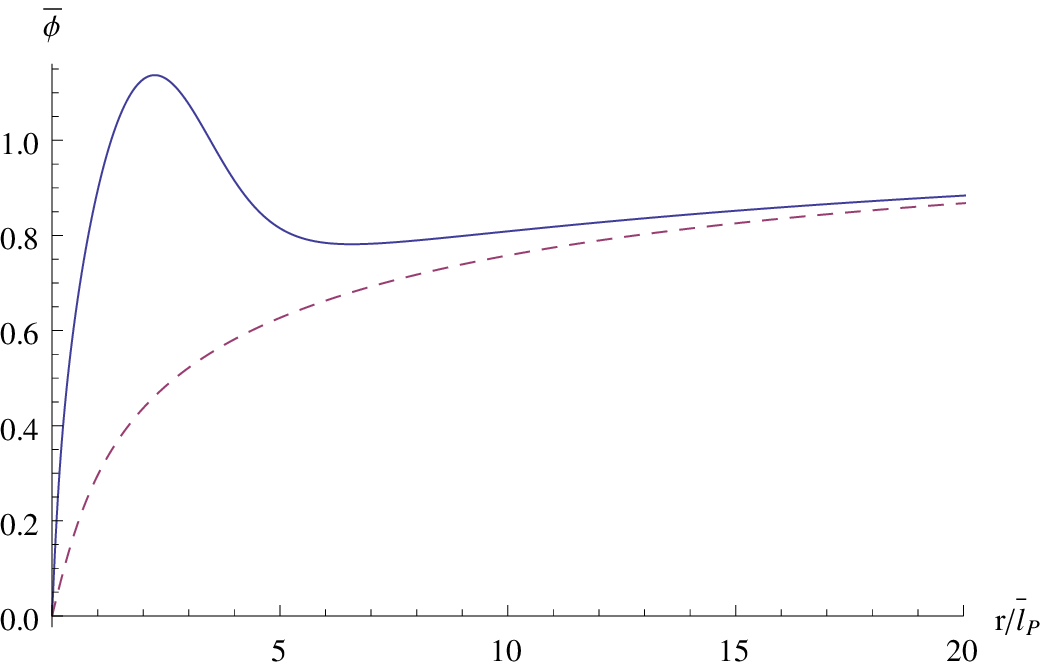}\\
\includegraphics[scale=0.55]{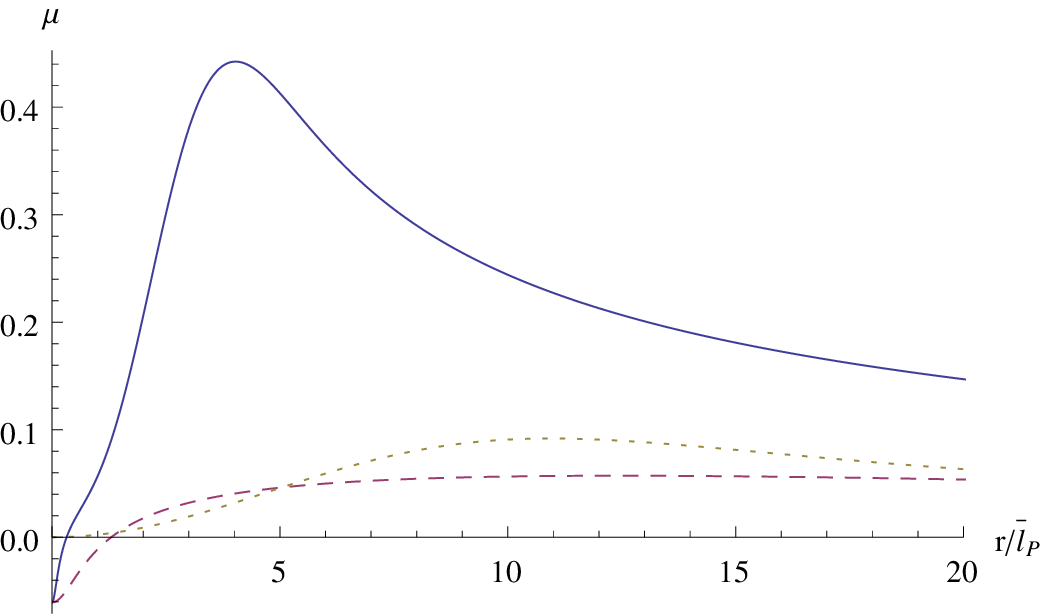}\qquad
\includegraphics[scale=0.55]{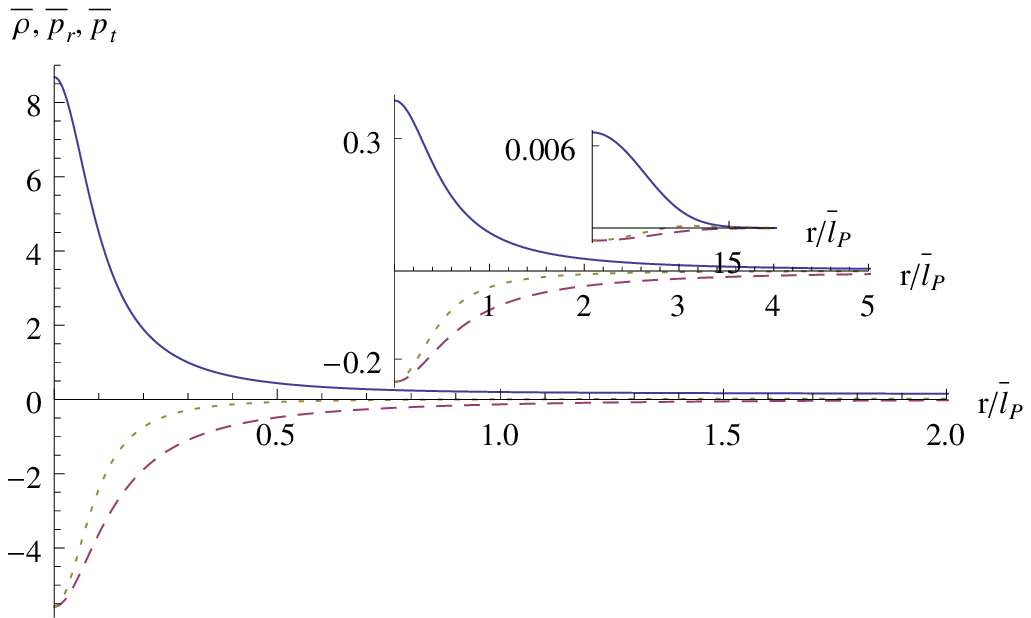}
\caption{Upper left: The boson star field alone (dashed) and in the presence of a global monopole (solid). Upper right: The global monopole field alone (dashed) and in the presence of a boson star (solid). Lower left: Compactness for the boson star alone (dotted), a global monopole alone (dashed) and a combined system (solid). Lower right: Energy density (solid)  of  the / combined system  / global monopole (inset) / boson star (smaller inset); radial pressure (dashed) and transversal pressure (dotted). $\sigma_0=0.05$,
$\lambda_{\rm BS}=0$, $\xi_{\rm BS}=-4$, $\lambda_{\rm GM}=0.1$,
$\Delta=0.08$, $\xi_{\rm GM}=5$.} \label{poljeBSGMa}
\end{figure}

It is a well known result that global monopoles in a very simplified model in which the field profile is modeled with a $\Theta$-function, exhibit a de Sitter interior ($p_{r,t}(0)=-\rho(0)$) and a Schwarzschild exterior with a deficit solid angle $1-\Delta-2GM/r$. The deficit solid angle is $\Delta=8\pi G_N\phi_0^2$ and denotes the scale of the symmetry breaking; $\phi_0$ is the vacuum expectation value of the field (minimum of the potential). Due to the de  Sitter core, global monopoles are gravitationally repulsive. Introducing nonminimal coupling, the deficit solid angle is modified $ \tilde\Delta=\Delta/(1+\xi\Delta)$, effective force could be locally attractive and bound circular orbits  around monopole could exist.

We show that a repulsive monopole stabilizes an attractive
boson star and the resulting configuration exhibits large energy density,
large (and negative) principal pressures, large compactness, large effective potential,
large local forces, and yet exhibit no event horizon (see Fig.~\ref{poljeBSGMa}). As such a composite
system of a boson star and a global monopole represents a convincing microscopic
candidate for a black hole mimicker.

\pagebreak

\end{document}